\documentclass[preprint2]{aastex}

\newcommand{\water}{H$_2$O}
\newcommand{\cotwo}{CO$_2$}
\newcommand{\methane}{CH$_4$}
\newcommand{\kms}{km\,s$^{-1}$}
\newcommand{\kp}{$K_\mathrm{P}$}
\newcommand{\kpval}{$(194^{+19}_{-41})$}

\newcommand{\rstar}{$R_\mathrm{S}$}
\newcommand{\rplan}{$R_\mathrm{P}$}

\newcommand{\prot}{$P_\mathrm{rot}$}
\newcommand{\protval}{$(1.7^{+2.9}_{-0.4})$}
\newcommand{\vrot}{$v_\mathrm{rot}$}
\newcommand{\vrotval}{$(3.4^{+1.3}_{-2.1})$}
\newcommand{\veq}{$v_\mathrm{eq}$}
\newcommand{\vsini}{$v\sin i_\star$}
\newcommand{\vrest}{$V_\mathrm{rest}$}
\newcommand{\vrestval}{$(-1.7^{+1.1}_{-1.2})$}
\newcommand{\pname}{\objectname{HD 189733\,b}}
\newcommand{\sname}{\objectname{HD 189733}}

\shorttitle{The synchronous rotation of HD~189733~b}
\shortauthors{Brogi et al.}

\begin{document}

\title{Rotation and winds of exoplanet HD~189733~b \\
    measured with high-dispersion transmission spectroscopy}


\author{M. Brogi\altaffilmark{1,2}, R. J. de Kok\altaffilmark{3,4}, S. Albrecht\altaffilmark{5}, I. A. G. Snellen\altaffilmark{3}, J. L. Birkby\altaffilmark{6,7} and H. Schwarz\altaffilmark{3}}
\email{matteo.brogi@colorado.edu}



\altaffiltext{1}{Center for Astrophysics and Space Astronomy, University of Colorado at Boulder, Boulder CO 80309, USA}
\altaffiltext{2}{NASA Hubble Fellow}
\altaffiltext{3}{Leiden Observatory, Leiden University, 2333CA Leiden, NL}
\altaffiltext{4}{SRON, Netherlands Institute for Space Research, Sorbonnelaan 2, 3584CA Utrecht, NL} 
\altaffiltext{5}{Stellar Astrophysics Centre, Department of Physics and Astronomy, Aarhus University, DK-8000 Aarhus C, DK}
\altaffiltext{6}{Harvard-Smithsonian Center for Astrophysics, Cambridge MA 02138, USA}
\altaffiltext{7}{NASA Sagan Fellow}

\begin{abstract}
Giant exoplanets orbiting very close to their parent star (hot Jupiters) are subject to tidal forces expected to synchronize their rotational and orbital periods on short timescales (tidal locking). However, spin rotation has never been measured directly for hot Jupiters. Furthermore, their atmospheres can show equatorial super-rotation via strong eastward jet streams, and/or high-altitude winds flowing from the day- to the night-side hemisphere. Planet rotation and atmospheric circulation broaden and distort the planet spectral lines to an extent that is detectable with measurements at high spectral resolution. We observed a transit of the hot Jupiter \pname\ around 2.3~\micron\ and at a spectral resolution of $R\sim10^5$ with  CRIRES at the ESO Very Large Telescope. After correcting for the stellar absorption lines and their distortion during transit (the Rossiter-McLaughlin effect), we detect the absorption of carbon monoxide and water vapor in the planet transmission spectrum by cross-correlating with model spectra. The signal is maximized (7.6$\sigma$) for a planet rotational velocity of \vrotval~\kms, corresponding to a rotational period of \protval~days. This is consistent with the planet orbital period of 2.2~days and therefore with tidal locking. We find that the rotation of HD~189733 b is longer than 1 day (3$\sigma$). The data only marginally (1.5$\sigma$) prefer models with rotation versus models without rotation. We measure a small day- to night-side wind speed of \vrestval~\kms. Compared to the recent detection of sodium blue-shifted by $(8 \pm 2)$~\kms, this likely implies a strong vertical wind shear between the pressures probed by near-infrared and optical transmission spectroscopy. 
\end{abstract}




\section{Introduction}\label{intro}

Giant exoplanets orbiting very close to their parent star (hot Jupiters) are thought to become tidally locked on timescales of 0.1-100 Myr \citep{ras96, mar97}, much shorter than the typical age of the observed systems. This configuration leads to extreme temperature contrasts between the permanently irradiated hemisphere and the night-side. The presence of an atmosphere can partially mitigate the temperature contrast, by recirculating the incident stellar energy from the day- to the night-side through winds. Modeling and observing the mechanisms driving the heat redistribution in exoplanet atmospheres is important to understand their global energy balance. 

One common outcome of both numerical \citep[e.g.,][]{sho09, rau10, hen11} and analytical \citep{sho11} simulations is that relatively deep in the planet atmosphere (at pressures of $\sim$1 bar) giant planets develop equatorial jet winds streaming eastward (e.g., following the planet rotation). Since those pressures are representative of the planet photosphere and therefore of the broad-band planet thermal emission, this deep circulation shifts the hottest point in the planet atmosphere away from the sub-stellar point. Higher-up in the atmosphere (at pressures $\le1$ mbar), winds predominantly flow from the day to the night side, crossing the planet terminator at all latitudes. Enriching this basic picture, high-altitude winds can be damped by magnetic drag due to the interaction of the partially-ionized atmosphere with a possible planetary magnetic field \citep[e.g.,][]{per10}. As a consequence, wind speeds and the relative importance of equatorial jets and global day- to night-side circulation are expected to vary broadly from planet to planet. Finally, the atmosphere and the planet interior do not always reach perfect tidal synchronization \citep{sho02}, meaning that global atmospheric super-rotation or sub-rotation are also possible.

Observational constraints on the energy recirculation of hot-Jupiter atmospheres have been placed by broadband observations of secondary eclipses and thermal phase curves \citep[see][for a review]{cow11}. Some work indeed confirms the presence of eastward displacements in the maximum of the thermal emission \citep{cro10, knu12, ste14}, which is in line with the above theoretical predictions. Other work either lacks the continuous phase coverage for assessing such a trend \citep{har06, cow07}, or show no evidence for a displacement \citep{knu09a, cro12}. Optical phase curves have been obtained mostly thanks to the {\sl Kepler} mission, which resulted in additional constraints on exoplanet albedo and winds \citep{est13, shp15}.

In recent years, high-resolution spectroscopy has enabled robust constraints on fundamental parameters of exoplanet atmospheres. Due to their unique fingerprint at high dispersion, molecular species can be robustly identified via line matching, in contrast to the potential ambiguities in the interpretation of planet spectra observed at low-resolution or with sparse photometric bands. In addition, high-dispersion spectroscopy is sensitive to the planet orbital motion. This offers an effective way to disentangle the Doppler-shifted planet spectrum from the contamination of our own atmosphere, which is instead essentially static in wavelength. For non-transiting planets the detection of the planet motion translates into a measurement of the actual planet mass (rather than an upper limit) and orbital inclination.

High-dispersion observations with CRIRES at the VLT led to the first measurement of the radial velocity of an exoplanet through the detection of CO in the transmission spectrum of HD~209458~b \citep{sne10}, and to the first detection of the atmosphere of a non-transiting planet \citep[$\tau$ Bo\"otis~b,][]{bro12, rod12}. These studies were conducted around 2.3\,\micron\ and targeted a ro-vibrational band of carbon monoxide, which is expected to be abundant in hot planet atmospheres and has a well-known and regularly-spaced sequence of lines. These factors make it one of the most favorable species for detection. More recently, \citet{bir13} and \citet{lock14} demonstrated the feasibility of high-resolution observations in the $L$-band and detected \water\ in the atmospheres of a transiting (\pname) and a non-transiting ($\tau$ Bo\"otis~b) planet. Water vapor has a much more complex high-resolution signature than carbon monoxide and the telluric absorption at 3.2-3.5~\micron\ is more severe, making it a good test for the robustness of these observational methods. It is also worth noting that the detection of \citet{lock14} was obtained by utilizing NIRSPEC at Keck, at only 1/4 of the resolving power of CRIRES (25,000 instead of 100,000), but with a wider wavelength range. This demonstrates that spectral resolution can be partially traded for spectral coverage for this type of observation. Finally, high-resolution spectroscopy was recently utilized to provide a first estimate of the C/O ratio in the atmosphere of HD~179949~b \citep{bro14}. The latter parameter is potentially related to the formation history of planets \citep{obe11} and to the presence of thermal inversion layers in their atmospheres \citep{mad12}. In this work we focus on the additional capabilities of high-dispersion observations to directly detect the Doppler signature of rotation and atmospheric circulation from a transiting exoplanet. Already \citet{sne10} reported a marginal detection of winds flowing from the day- to the night-side of HD~209458~b, deduced from the measured ($-2\pm1$)~\kms\ blue-shift in the total cross-correlation signal. This result was further discussed by \citet{mil12}, \citet{sho13}, and \citet{kem14}. They linked the theoretical expectations based on atmospheric circulation models to the observable Doppler signature at high spectral resolution. It is worth noting that the alternative interpretation of the $-2$~\kms\ shift as due to orbital eccentricity \citep{mon11} is excluded by the latest refinement of the planet orbital parameters \citep{cro12, sho13}. At optical wavelengths, \citet{wyt15} also claimed high-altitude winds in the atmosphere of HD~189733~b by measuring a blue-shift in the planet sodium absorption of $(-8\pm2)$~\kms. More recently, \citet{sne14} inferred the fast rotational rate of the young, directly-imaged, giant planet $\beta$~Pic~b by measuring the broadening of the planet spectral features. In this case the rotational period of only $(8\pm1)$~hours produced a broadening of 25~\kms, well resolved at the spectral resolution of CRIRES (FWHM~$\simeq 3$~\kms).

Here we search for the Doppler signature of synchronous rotation and atmospheric circulation in the high-resolution transmission spectrum of \objectname{HD 189733 b} \citep{bou05}. This is one of the best-studied exoplanets to date, thanks to the apparent brightness of the host star (K = 5.54 mag). This richness of information translates into observational evidence for rather unique features, such as atmospheric escape \citep{lec10, jen12} and the prominence of haze \citep{pon08, sin11} masking most of the spectral features in the optical transmission spectrum, except for atomic sodium \citep{red08, hui12, wyt15} and potassium \citep{pon13}.
This planet also has some constraint on energy recirculation and day/night side contrasts from multi-wavelength infrared phase curves \citep{knu07, knu09b, knu12}. One of the most distinctive features of the phase curve is a displacement of the planet hot-spot towards the east, and a low day to night side temperature contrast at shorter infrared wavelengths, indicating an efficient heat recirculation between the two hemispheres. Low-resolution observations of the near-infrared spectrum of \pname\ point towards the presence of water and possibly carbon monoxide \citep{bar07, bea08, cha08, des09}. At high-resolution, detections of carbon monoxide in the $K$-band \citep{rem13} and \water\ in the $L$-band \citep{bir13} were reported from dayside observations.

In this work we analyze near-infrared, high-resolution spectra of HD~189733~b observed just before and during the planet transit. Beside searching for the planet absorption lines due to CO, \cotwo, \water, and \methane, we aim to detect the spectroscopic signatures of rotation and atmospheric circulation. In Section~\ref{observations} we describe the telescope, instrument, and observations. In Section~\ref{analysis} we present our data analysis, with a particular focus on the subtraction of the stellar spectrum to avoid contamination from the stellar CO lines (Section~\ref{rm_star}). In Section~\ref{extraction} we illustrate how the models for the planet atmosphere are computed and the signal is extracted. The detected signal is presented in Section~\ref{results} and the implications for the atmosphere of \pname\ are discussed in Section~\ref{discussion}, together with some general advice for future high-dispersion observations of exoplanets.

\section{Observations}\label{observations}

We observed one transit of \objectname{HD 189733 b} with the CRyogenic Infra-Red Echelle Spectrograph \citep[CRIRES,][]{kae04} at the ESO Very Large Telescope (VLT) facility. The spectrograph is mounted at the Nasmyth-A focus of the Antu Unit. We opted for the highest available spectral resolution ($R$~=~100,000) by observing through the 0$\arcsec$.2 slit, and we selected the standard wavelength setting with $\lambda_\mathrm{ref}=2325.2$~nm, approximately covering the range 2287.5-2345.4~nm. The spectra are imaged on four detectors having $1024\times512$ pixels in the spectral and spatial direction respectively. Gaps of 282, 278, and 275 pixels in the spectral direction are due to the physical separation of each detector. The throughput of CRIRES was maximized by observing in conjunction with the Multi Application Curvature Adaptive Optic system \citep[MACAO,][]{ars03}. For accurate background subtraction, we nodded by 10\arcsec\ along the slit with no jitter, according to an ABBA sequence.

The transit of HD~189733~b was observed with the DDT proposal 289.C-5030 on 2012, July 30. The observations started at 1:10~UT and were stopped at 3:14~UT due to pointing restrictions, corresponding to a range in planet orbital phase $\varphi$ of 0.977-0.017. This was sufficient to observe a full transit of the planet, with the first contact happening at 1:25~UT and the fourth contact during the last exposure. The first spectra were taken at airmass greater than 2 with a seeing of 1.2-1.4\arcsec. At 1:22~UT the seeing abruptly dropped below 1\arcsec, resulting in significantly higher signal-to-noise ratio due to better performances of the AO system. With 60 seconds of exposure time per spectrum, we collected 45 pairs of AB or BA spectra, of which 39 are during the transit of HD~189733~b.


\section{Data analysis}\label{analysis}

We utilized the CRIRES pipeline version 2.1.3 for the basic calibration and the extraction of the one-dimensional spectra. On each frame, dark-subtraction and flat-field correction were performed by utilizing the standard calibration frames observed the morning after the transit. An additional non-linearity correction was applied on the science and calibration frames by using the appropriate set of archival data.
Each pair of AB or BA frames was subsequently combined in order to subtract the background, and the one-dimensional spectrum was obtained by optimal extraction \citep{hor86}. 

A difference in gain between odd and even columns is known to affect detectors 1 and 4 (odd-even effect). A set of calibration files is provided with the pipeline to correct for this effect. However, in all our previous work with CRIRES we noticed that these files are able to correct the odd-even effect in detector \#1, but not in detector \#4. We therefore discard the latter for the rest of this work.

The remainder of the data analysis was performed via custom-built procedures in IDL. For each of the CRIRES detectors, the extracted 1-D spectra were organized in two-dimensional arrays containing pixel number (wavelength) and frame number (time/phase) on the horizontal and vertical axes, respectively. 

Bad-pixels due to cosmic rays and detector cosmetics were corrected as in \citet{bro14}, by spline interpolating across isolated pixels or bad columns in the data matrix, by linearly interpolating across 2-3 consecutive bad pixels in the spectral direction, and by masking larger groups of bad pixels.


\begin{figure}[ht]
\centering
\plotone{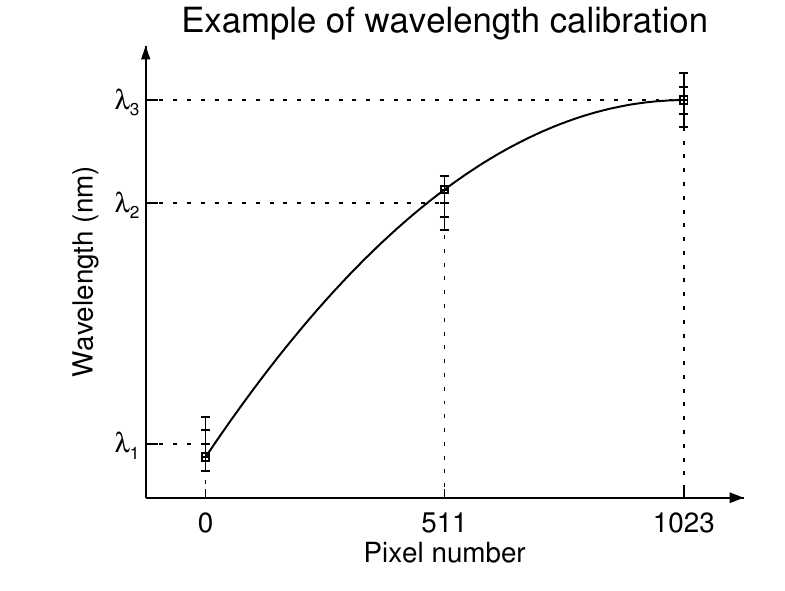}
\caption{Our iterative algorithm for wavelength calibration. We start with a triplet ($\lambda_1,\lambda_2,\lambda_3$) of guess values for the wavelengths associated to pixels $x=(0,511,1023)$. We explore a grid of 5 equally-spaced points around each guess value, resulting in 125 possible combinations. Each combination has a quadratic solution associated (the parabola passing through the three points, solid line). The best solution is determined as described in Section~\ref{align}, and the triplet ($\lambda_1,\lambda_2,\lambda_3$) is updated accordingly. The final solution is found iteratively, by refining the grid step by a factor of 5 at each iteration.}
\label{fig_cal}
\end{figure}


\subsection{Alignment of spectra and wavelength calibration}\label{align}

In order to implement our strategy for removing telluric absorption (see Section~\ref{rm_tell}), the spectra need to be aligned to a common wavelength scale. In previous work \citep{bro12, bro14}, this task was performed by measuring the centroid position of the telluric lines, by fitting their displacement with respect to a reference spectrum with a linear function in pixel position, and by shifting the spectra back to the reference by spline interpolation. However, the star \sname\ shows absorption lines in the observed wavelength range, subject to a changing Doppler shift due to the changing barycentric velocity of the observer and the stellar radial velocity. Since these effects could potentially impact the centroid determination, in this work we did not measure the centroids of telluric lines, but we utilized the cross correlation between the data and a template to align and determine the wavelength solution of the data. The template spectrum contains telluric and stellar lines. The telluric model was computed via the ESO Sky Calculator\footnote{https://www.eso.org/observing/etc/bin/gen/form?\\INS.MODE=swspectr+INS.NAME=SKYCALC} with parameters automatically set by the software after selecting the observing night and the time of our observations. This means that we computed a telluric model for each observed spectrum. The stellar template only included CO lines. Their rest wavelengths $\lambda_\mathrm{rest}$ were extracted from the HITRAN database and shifted to the observed wavelengths $\lambda_\mathrm{shift}$ by applying
\begin{eqnarray}
\lambda_\mathrm{shift} & = & \lambda_\mathrm{rest}(1 + v_\mathrm{S}/c) \\
v_\mathrm{S} & = & v_\mathrm{bary} + v_\mathrm{sys} + K_\mathrm{S}\sin[2\pi(\varphi+0.5)],
\end{eqnarray}
where $v_\mathrm{S}$ is the stellar radial velocity, $c$ is the speed of light, $v_\mathrm{bary}$ is the barycentric velocity of the observer, $v_\mathrm{sys}$ is the systemic velocity, $K_\mathrm{S}$ the stellar radial-velocity semi-amplitude, and $\varphi$ the planet orbital phase ($\varphi=0$ at mid-transit). The relevant parameters of the system \sname\ are listed in Table~\ref{tab_syspar}. The CO line depth was taken to be equal to the absorption coefficients for a temperature of 5000~K, multiplied by 10$^{22}$, which was found to be a good visual match to the data. Finally, the line profile was set to be a Gaussian with FWHM~=~3~\kms, which corresponds to a spectral resolution of $R=10^5$ and unresolved line profiles. Although in Section~\ref{rm_star} we also need to account for changes in instrumental and stellar line profiles, we found that the alignment algorithm does not require this level of detail. For each observed spectrum, the global template spectrum was obtained by multiplying the telluric and stellar template constructed as above. 

We determined the wavelength solution of each spectrum and each detector by calculating a grid of trial quadratic solutions, which is appropriate for CRIRES data according to the manual\footnote{\url{www.eso.org/sci/facilities/paranal/instruments/crires/doc.html}}.
As shown in Figure~\ref{fig_cal}, each quadratic solution is given by the parabola passing through the wavelengths of the first, middle, and last pixels of the detector ($\lambda_1,\lambda_2,\lambda_3$). The starting values were given by the default (pixel, wavelength) solution of the CRIRES pipeline. The parameter space around these three values was sampled using five equally-spaced steps, centered around the initial estimates. The initial step size was chosen to be 0.2~nm, big enough to encompass all possible solutions. Each of the resulting 125 grid points defines a trial wavelength solution for the data, to which we interpolated the global template. We then computed the cross correlation with the data, selected the best solution based on the highest cross-correlation, and updated the triplet of wavelengths ($\lambda_1,\lambda_2,\lambda_3$). We refined the step size by a factor of 5 (0.04 nm) and repeated the computation. After 4 iterations, we adopted the best-fitting solution as wavelength solution for our spectrum. In the absence of noise this would give us a precision of $3.2\times10^{-4}$~nm (0.03 pixels) in the wavelength solution.

We finally aligned all spectra to a common wavelength scale, which we computed by averaging the wavelength solution of all spectra. The alignment was performed by spline-interpolating each spectrum to the new wavelength scale.


\subsection{Fit and removal of the stellar spectrum}\label{rm_star}
During a large portion of the transit, star and planet have a similar radial velocity (to within few \kms), meaning that spectral features present in both the stellar and the planet spectrum are almost superimposed. This is the case of the CO ladder around 2.3~$\mu$m, which is a distinctive feature of both the planet and the stellar spectrum.
Stellar CO lines have depths of 10-20\% with respect to the continuum, while the typical depth of the planet spectral lines is expected to be 100-200 times smaller.

\begin{figure}[!ht]
\centering
\plotone{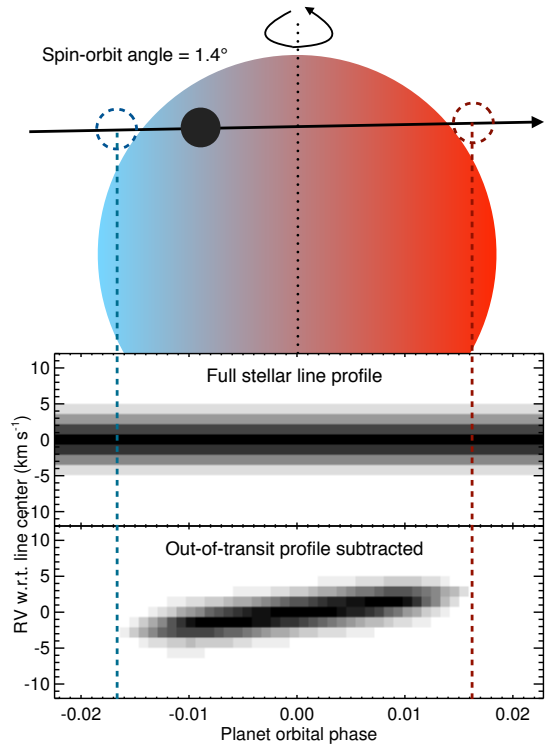}
\caption{Doppler signature caused by the Rossiter-McLaughlin effect. The rotating stellar disk and the path crossed by the transiting planet are illustrated in the top diagram, including the relative angle of 1.4$^\circ$ between the stellar spin and the normal to the orbit. The transiting planet occults varying portions of the rotating stellar disk (from the blue-shifted to the red-shifted hemisphere) inducing a distortion in the stellar line profile. This is not visible in the middle panel, where we show a full stellar line profile simulated as in Section~\ref{rm_star}. However, this distortion becomes evident when subtracting the out-of-transit stellar line profile from the in-transit distorted profile (bottom panel). These residuals mimic the Doppler signature of a transiting exoplanet at a relatively low orbital velocity, causing spurious CO absorption (see Figure~\ref{fig_co_res}) if not modeled and removed from the data. The grey scale is linear in the middle panel, linear and inverted in the bottom panel.}
\label{fig_rm_effect}
\end{figure}

The stellar spectrum cannot be normalized in the same way as the telluric spectrum (see Section~\ref{rm_tell}), because the stellar absorption lines are distorted by the Rossiter-McLaughlin (RM) effect \citep{ros24, mcl24, que00}. During different phases of a transit, particular parts of the rotating stellar photosphere are blocked from view, leading to a deformation of the stellar absorption lines. This deformation and its progress during transit depends on the exact geometry of the transit, which is governed by the projected stellar rotation (\vsini), the impact parameter, and the projected angle between the stellar and orbital angular momenta (see Figure~\ref{fig_rm_effect}). The RM effect is often used to measure the projected obliquities \citep[e.g.][]{heb08, alb12} and for the case of \pname\ good alignment was found \citep{win06,tri09,cam10}.

For our current investigation the RM effect does however present a nuisance as it causes spurious time-varying signal during transit. For slowly rotating stars like HD\,189733, the magnitude of the deformation on the absorption lines roughly scales with (\rplan/\rstar)$^2$. The stellar absorption lines are 100-200 times stronger than the planetary CO lines. Therefore RM effect is roughly of the same order of magnitude as the expected signal from the planetary CO absorption and therefore needs to be treated properly. Although the RM effect does not contaminate the spectra of \water, \methane, and \cotwo, we expect CO to have an important contribution to the planet spectrum as well, based on previous high-resolution dayside observations of \pname\ \citep{rem13}. We therefore decided to model the distortion of the stellar lines during transit, fit the stellar absorption lines in our data, and remove the resulting stellar spectrum from these data, as outlined below.

For computing the RM effect we used the numerical model by \citet{alb07}, already applied to eclipsing binary stars \citep[e.g.,][]{alb14} and transiting planets \citep{alb13}. In short this numerical model simulates a rotating absorption line kernel assuming solid-body rotation, quadratic limb darkening (Equation~\ref{eq_starint}), and the macro-turbulence model by \citet{gra84}. Solar-like convective blueshift is also included \citep{shp11}. The micro turbulence is modeled by a simple Gaussian. The resulting distorted stellar absorption lines, appropriate for a particular transit phase, are then used to convolve the narrow stellar model spectra in the fitting algorithm described below.

For this RM model we assumed good projected alignment, an orbital inclination of $85.71^\circ$, and solar like blueshift, but with a reduced amplitude of $200$~m\,s$^{-1}$, compared to the solar value of $500$~m\,s$^{-1}$. The limb darkening parameters had been set to $u_1=0.077$ and $u_2 = 0.311$, appropriate for a star with $T_\mathrm{eff}=5000$\,K, $\log\,g=4.5$\,cm\,s$^{-2}$, and solar metallicity, observed in the $K$-band \citep{cla11}. The projected rotation speed \vsini, macro turbulence, planet/star radii ratio (\rplan/\rstar), and the micro turbulence have been initially left as free parameters and have been adjusted throughout the procedure described below. However, since our fit is scarcely influenced by \vsini\ and \rplan/\rstar\ when varied within their literature uncertainties, we fixed these two parameters to the values listed in Table~\ref{tab_syspar}. This allowed us to sensibly reduce the computational time.

\begin{figure}[!th]
\centering
\plotone{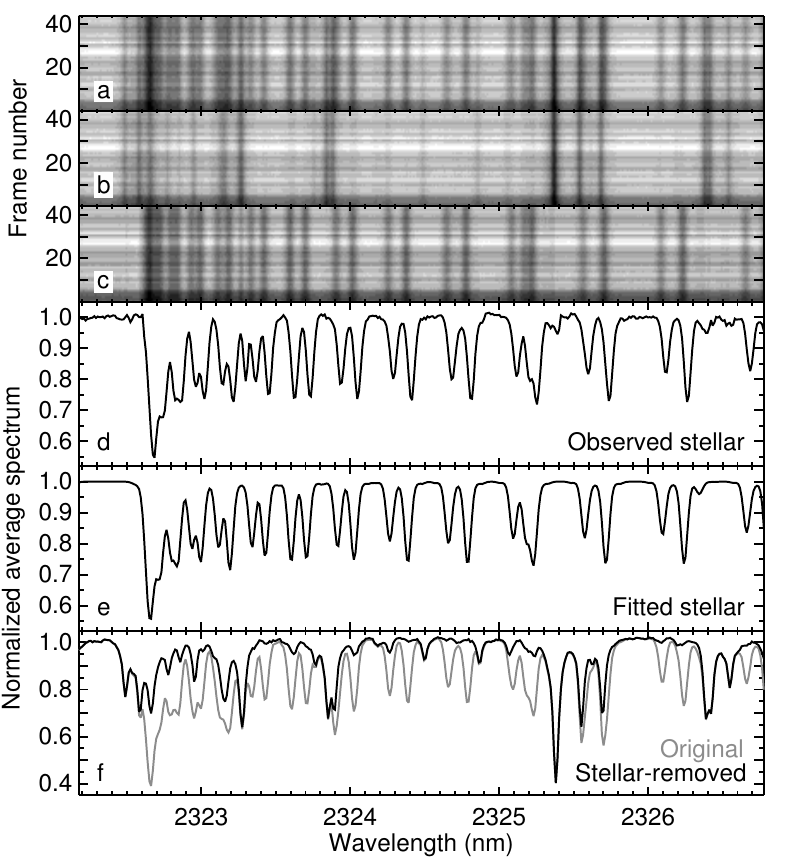}
\caption{Removal of the fitted stellar spectrum from our data. The six panels (a-f) correspond to six steps of the iterative fitting process described in Section~\ref{rm_star}. The top three panels show the sequence of 45 spectra of HD~189733 around 2324.4 nm. Panel (a) includes all spectral components (stellar and telluric), while in panels (b) and (c) the best-fitting stellar and telluric spectra were removed. By averaging the out-of-transit spectra of panel (c) we obtain the average stellar spectrum (panel d) and its best fit (panel e). After iterating the algorithm a few times, the final best-fitting stellar model is removed from the full spectrum. Panel (f) shows an average of the 45 observed spectra before (in grey) and after (in black) the stellar removal.}
\label{fig_rm_star}
\end{figure}

We designed an iterative procedure to fit simultaneously for the instrumental profile (IP) of CRIRES, for the depth and shape of stellar CO lines, and for residual non-CO stellar lines. The fitting algorithm is outlined below and illustrated in Figure~\ref{fig_rm_star}. In the following, we refer to the original spectra as the data after alignment, containing telluric and stellar absorption lines (panel a).

\begin{enumerate}
\item We divide the original spectra through the Doppler-shifted stellar template (see Section~\ref{align}) utilized for the alignment (panel b);
\item We obtain an initial guess for the IP by applying the singular value decomposition algorithm described in \citet{ruc99} and implemented in \citet{sne14} on the residual spectra after step 1. As narrow template, we use the same telluric model from the ESO Sky Calculator as in Section~\ref{align}. We note a negligible influence of the kernel size chosen to determine the IP on the final result, for sizes between 9 and 21 CRIRES pixels, where a pixel is 1.5~\kms. We choose a kernel size of 19 pixels, corresponding to 4 FWHM of a theoretical Gaussian IP at the spectral resolution of 100,000;
\item We convolve the ESO telluric model with the guess IP and we divide the original data through it, obtaining the stellar spectrum plus residual noise from imperfect telluric subtraction (panel c);
\item We obtain a master stellar spectrum by shifting the out-of-transit spectra to the stellar rest frame and co-adding them (panel d);
\item We fit the stellar CO lines in the master spectrum by shifting the synthetic stellar line profiles computed above to their Doppler-shifted wavelengths ($\lambda_\mathrm{shift}$ in Section~\ref{align}), by convolving them with the IP measured during step 2, and by fitting the correct scaling factor by linear regression. We find the best-fitting micro-turbulent and macro-turbulent velocities by minimizing the $\chi^2$ of the residuals;
\item We fit residual non-CO stellar lines with Gaussian or Lorentzian profiles. These lines are not supposed to produce spurious planetary signal, therefore a precise fitting is not required. The total fitted stellar spectrum (CO and non-CO lines) is shown in Figure~\ref{fig_rm_star} (panel e);
\item We repeat steps 1-6 iteratively until the IP, the minimum $\chi^2$, and the best-fit parameters converge. This is typically achieved after 3 iterations. At point 1, we substitute the stellar template utilized in Section~\ref{align} with the fitted stellar spectrum after steps 5 and 6.
\end{enumerate}
We divide the original data through the fitted stellar model, which includes the non-CO lines fitted during step 6 above, and the modeled RM effect for the CO lines and in-transit spectra. The best-fit stellar line profile is convolved with the IP measured as above, shifted to the observed frequency accounting for the barycentric velocity of the observer and the stellar radial velocity, and scaled by the measured amplitude.
An example of our data before and after the removal of the stellar spectrum is shown in Figure~\ref{fig_rm_star} (panel f).

\subsection{Effect of star spots}
Spots on a rotating stellar photosphere are capable of distorting the stellar line profile depending on their longitude and latitude. Since we do not model this effect, our fit might be partially incorrect, leading to small residuals within $-$\vsini\ and +\vsini\ from the stellar radial velocity. 
We demonstrate below that, due to the nature and amplitude of the line distortion caused by star spots, their effect is negligible for this study. 

In the first place it is useful to estimate the amplitude of these distortions. \citet{win07} estimated a spot coverage of 1\% at any moment for the star HD~189733, based on optical multi-band photometric variability. The spot temperature was estimated by \citet{pon08} to be 500-1000~K colder than the photosphere ($T_\mathrm{eff}\simeq5000$~K). We adopted $\Delta T = 750$~K and consequently a spot temperature of $T_\mathrm{spot} = 4250$~K. With these values, the maximum effect of spots at 2.3~\micron\ is $2.6\times 10^{-3}$ relative to the depth of a stellar line, if the continuum and the spots are approximated with black bodies emitting at temperatures $T_\mathrm{eff}$ and  $T_\mathrm{spot}$, respectively. For an average CO line depth of 15~\%, these residuals have amplitudes of $\sim4\times10^{-4}$ compared to the stellar continuum, which is at least $10\times$ smaller than the weaker CO lines in the planet transmission spectrum.

Furthermore, in Doppler space the nature of this spurious signal is substantially different from the planet absorption, and largely canceled prior to cross-correlation. \citet{win07} demonstrated that HD 189733 is a slow rotator with a rotational period close to 13 days. Thus any configuration of spots can be considered static during one single transit observation (less than 2 hours). Compared to the observer, the star moves in radial velocity by only 0.2~\kms\ during the transit (mostly due to the change in barycentric velocity), which is only 1/15 of the full width half maximum of the CRIRES IP. This means that any distortion of the stellar lines caused by unocculted spots is static in both radial velocity and time. In this case our data analysis designed to remove telluric lines (Section~\ref{rm_tell}) will also re-normalize residuals from uncorrected star spots, further dampening the already small signal.

However, we know from past transit observations \citep{pon07, pon08} that HD~189733~b usually crosses some spots during transit. In this case, the distortion of the stellar line suddenly disappears (when the spot is occulted) and reappears, generating a pulse radial velocity signal in the data, uncorrected by our analysis. The duration of this pulse is dependent on the size of the crossed spot (or region of spots). Although probably unrealistic, we discuss the worst-case scenario in which all spots are clustered in a region of radius 0.1~\rstar\ (based on the ~1\% coverage and spots significantly darker than the photosphere). For an impact parameter of $b=0.66$, the length of the chord crossed by the planet is 1.5 stellar radii. With 39 frames observed in transit, the planet crosses 0.038 stellar radii per frame. This unrealistically big group of spots would therefore affect 2.6 frames of our data. For a spot signal 1/10 of the planet signal, occulted star spots can only contaminate 8\% of the detected CO signal, or $\sim0.3\sigma$ based on the detection presented in Section~\ref{onegas}. We therefore conclude that - at least in the case of HD 189733 b - the contamination due to uncorrected star spots is negligible.


\subsection{Removal of telluric lines}\label{rm_tell}
Telluric lines change in depth during the observations, but not in position. Therefore, the telluric spectrum is most effectively removed by modeling the flux of each data column (i.e., each wavelength channel) in time. The main cause of telluric variation is the change in geometric airmass. Changes in the IP (due to, e.g., in-slit guiding, seeing, and performances of the adaptive optics), as well as changes in the water vapor content in the Earth's atmosphere could also affect the flux in the core of telluric lines. 

We initially normalize each spectrum by its median continuum level. We then model the time-variation of the flux in each spectral channel via the simple functional form:
\begin{equation}
F_\lambda(t) = a_0 + a_1A(t) + a_2A^2(t),
\end{equation}
where $A$ is the geometric airmass, and we divide through the fit. The coefficients $a_0$-$a_2$ are determined by linear regression for each of the spectral channels in the data. We investigated more complex functional forms for $F_\lambda(t)$, but since the gain in total planet signal was marginal, we preferred to keep the most conservative and least invasive decorrelation. We divide each spectral channel by its variance, which is required because the data have been re-normalized and we therefore need to restore the initial S/N of each spectral channel through an additional weighting. Finally, we apply a high-pass filter to each spectrum by computing and dividing through a smoothed version (boxcar of width 64 pixels).

We note that the approach described above partially differs from previous work \citep{sne10, bro12, bro14}. There we preferred to fit a first-order airmass dependence, sample higher-order residuals directly from the data, and determine their linear combination that best matched $F_\lambda(t)$. This method is particularly effective when planet molecular lines have depths well below the level of the noise in the data. In this case, molecular lines in the transmission spectrum of \pname\ are expected to be as deep as $8\times10^{-4}$ relative to the stellar continuum, which is close to the typical noise levels of these data (average S/N = 270 per spectrum). Sampling the residuals from the data is therefore not advisable, because we would likely include and consequently remove part of the planet signal as well. 


\section{Extracting the planet signal}\label{extraction}

After removing the stellar and telluric lines as explained in Sections~\ref{rm_star} and \ref{rm_tell}, we extracted the planet signal by cross correlating the data with model spectra for the planet atmosphere. In this way we combined all the spectral lines of the planet in a single cross-correlation function (CCF), increasing the signal by a factor of $\sqrt{N}$, where $N$ is the number of strong molecular lines.

We computed the cross correlation with both narrow and broadened atmospheric models. The former are the output of our radiative transfer calculations described in Section~\ref{atm_models}, which do not account for atmospheric circulation or rotational broadening. The latter are the narrow models convolved with the broadening profiles described in Section~\ref{broad_models}. 


\subsection{Modeling the planet spectra}\label{atm_models}

The transmission spectrum of HD~189733~b was modeled by following the same prescriptions of \citet{bro14} concerning the choice of line lists for CO and \water\ \citep{rot10}, \methane\ \citep{rot09}, line profiles, and collision-induced absorption by hydrogen \citep{bor01, bor02}. The line list for \cotwo\ was instead taken from \citet{rot95}. We described the terminator atmosphere of the planet by assuming an average temperature-pressure ($T/p$) profile. The lower part of the planet atmosphere ($p>0.1$ bar) cannot be probed by transmission spectroscopy. Any chosen $T/p$ profile would essentially produce the same modeled transmission spectrum. Therefore, we chose to match the \citet{mad09} profile, which is derived from secondary-eclipse measurement of the planet, with the simplest possible sampling. For $p>1$~bar, an isothermal atmosphere at temperature 1750~K was adopted. Between 1 and 0.1~bar, temperature decreased to 1350~K. In the range $-2.5<\log(p)<-1.0$, we tested instead two opposite profiles, by making the temperature $T_2$ either increase to 1500~K or decrease to 500~K. For lower pressures (higher altitude), we chose an isothermal atmosphere at temperature $T_2$, extending up to $p=10^{-10}$~bar. We note that $T_2=1500~K$ corresponds to a temperature inversion, which is included in these models reflecting our very limited knowledge of the upper atmosphere of this planet. However, the inversion is so weak and so high in the atmosphere that it does not affect dayside spectra (i.e., it would not produce emission lines in the planet thermal spectrum). The only effect of $T_2$ on the transmission spectrum is to change the relative strength of weak and deep molecular lines by changing the scale height at high altitude. The radiative transfer was performed by computing the optical thickness for 40 altitude points along slant paths through the atmosphere, and then by integrating over altitude and the $2\pi$ radians of the planet limb to produce the transmission spectrum.

We considered the four main trace gases expected to be spectroscopically active in the observed wavelength range \citep{mad12}: CO, \water, CO$_2$, and \methane. These are also the gases for which detections have been claimed in the literature, although some of them are still under debate (see Section~\ref{intro}). We investigated molecular Volume Mixing Ratios (VMRs) of ($10^{-5},10^{-4},10^{-3}$) for carbon monoxide and water vapor, and VMR~=~($10^{-7},10^{-5}$) for carbon dioxide and methane. 
We first computed models with the four species as single trace gases, and we detected CO and \water\ (Section~\ref{onegas}). We then computed models with CO and \water\ mixed and determined the combination that maximized the planet signal (Section~\ref{best_model}). Examples of our modeled transmission spectra are shown in the left column of Figure~\ref{fig_moldet}.


\subsection{Modeling the planet broadening profile}\label{broad_models}

We modeled the broadened planet line profiles by assuming two basic patterns of atmospheric circulation: bulk planet rotation and equatorial super-rotation. A global day-to-night side flow does not need modeling, as it produces a blue-shifted cross-correlation signal that is measurable directly after shifting and co-adding the CCFs. 

\begin{figure}[!t]
\centering
\plotone{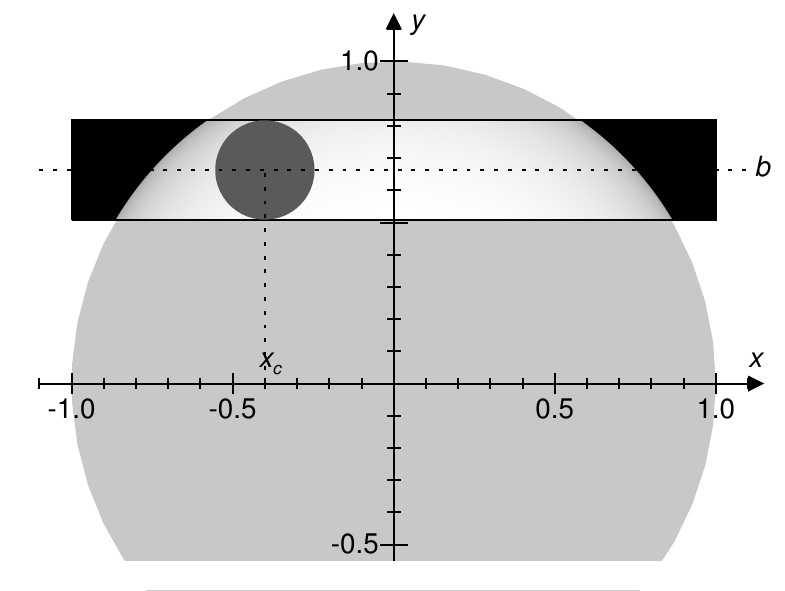}
\caption{Basic geometry of the model utilized for computing the planet broadening profiles. All quantities are expressed in units of stellar radius. The intensities of 100 slices of the stellar disk, corresponding to impact parameters between $b-R_\mathrm{P}/R_\mathrm{S}$ and $b+R_\mathrm{P}/R_\mathrm{S}$ (the boxed area in the figure), are computed based on a quadratic limb-darkening law (Equation~\ref{eq_starint}). The planet, centered in ($x_\mathrm{C},b$), is shifted along the $x$ direction pixel by pixel. At each position, the light curve is computed as explained in Section~\ref{broad_models}.}
\label{fig_model}
\end{figure}

We constructed a two-dimensional model for the planet and the star, which is represented in Figure~\ref{fig_model}. The literature values for all the relevant quantities utilized below are listed in Table~\ref{tab_syspar}. We chose the $x$ and $y$ directions along and perpendicular to the orbit, respectively. We assigned the coordinates $(x,y)=(0,0)$ to the center of the stellar disk and expressed all quantities in units of stellar radius, \rstar. 
The planet was modeled as a disk of radius $n_\mathrm{P}=50$ pixels. With this choice the step size is $dx=$\rplan/(50~\rstar)~$\simeq0.0031$, with \rplan\ being the planet radius. Each horizontal slice of the planet disk corresponds to an impact parameter $b_i$ ranging from $(b-R_\mathrm{P}/R_\mathrm{S})$ to $(b+R_\mathrm{P}/R_\mathrm{S})$, where $b$ is the known impact parameter of HD~189733~b. 
The planet center has coordinates $(x_\mathrm{C},b)$ corresponding to an orbital phase:
\begin{equation}
\varphi(x_\mathrm{C}) = \frac{1}{2\pi}\sin^{-1}\left(\frac{x_\mathrm{C}}{a}\right),\end{equation}
where $a$ is the semimajor axis of the orbit. For each $b_i$, the intensity of the stellar chord crossed by the planet during transit was modeled by assuming a quadratic limb-darkening law:
\begin{equation}\label{eq_starint}
I(\mu)=1 - u_1(1-\mu) - u_2(1-\mu)^2,
\end{equation}
with $\mu$ being the cosine of the angle between the line of sight and the normal to the stellar surface, and $u_1=0.077,u_2=0.311$ the same limb-darkening coefficients as in Section~\ref{rm_star}. Given the geometry of our problem, Equation~\ref{eq_starint} can be expressed as function of $(x,b_i)$ through the following:
\begin{equation}\label{eq_mu}
\mu = \sqrt{1 - x^2 - b_i^2}.
\end{equation}
Equation~\ref{eq_mu} is valid for $-\bar{x}_i<x<\bar{x}_i$, with $\bar{x}_i$ being the coordinate of the edge of the stellar disk:
\begin{equation}
\bar{x}_i = \sqrt{1 - b_i^2}.
\end{equation}
The stellar intensity $\mathcal{S}_i(x)$ for the slice $i$ is therefore:
\begin{equation}
\mathcal{S}_i(x) = \left\{ 
\begin{array}{ll}
I(x,b_i) & \mbox{ for }  |\,x\,| \le \bar{x}_i\\
0 & \mbox{ elsewhere}.  
\end{array}
\right.
\end{equation}
At the edges of the stellar and planet profiles, sub-pixel approximation is taken into account by multiplying each pixel values by their fractional occupancy.

We tested the precision of our model against the \citet{man02} model. We assumed a completely opaque planet disk (0 pixel values). The transit light curve was computed by shifting the planet array one pixel at a time along the $x$ direction, in the range $-1.2<x_\mathrm{C}<1.2$. At each position, the planet and stellar arrays were multiplied and summed across all pixels in order to obtain the flux. Finally, the light curve was normalized by the total stellar intensity $I_\mathrm{tot}$, which we computed analytically via the following integral:
\begin{equation}
I_\mathrm{tot} = \int_0^{2\pi}\int_{0}^{n_\mathrm{S}}{I(r)r~dr~d\theta} = 2\pi\int_{0}^{n_\mathrm{S}}{I(r)r~dr},
\end{equation}
where $n_\mathrm{S}=1/dx$ is the number of pixels corresponding to one stellar radius. In the above we have used polar coordinates ($r,\theta$) with origin in the center of the stellar disk. Consequently, $\mu=\sqrt{1-(r/n_\mathrm{S})^2}$, which leads to 
\begin{equation}
I_\mathrm{tot} = \pi\left(1-\frac{u_1}{3}-\frac{u_2}{6}\right)n_\mathrm{S}^2 \simeq 3.026\times10^5.
\end{equation}
We found that our normalized light curve is in agreement with the \citet{man02} models at the $3.5\times10^{-5}$ level in relative flux. Consequently, we added a planet atmospheric ring with the same radius as the planet and with a thickness of 1 pixel. This corresponds to $\sim1600$~km in physical units, matching the 5-10 scale heights typically probed by transmission spectroscopy \citep{mad14}. We assigned to each pixel of the ring a velocity consistent with a rigid-body rotation along an axis perpendicular to the orbital plane:
\begin{equation}
v_\mathrm{ring} = \frac{2\pi x_\mathrm{ring}R_\mathrm{S}}{P_\mathrm{rot}},
\end{equation}
where $P_\mathrm{rot}$ is the planet rotational period and $x_\mathrm{ring}$ the $x$-coordinate of each ring pixel in units of \rstar. An additional equatorial super-rotation is introduced by adding (or subtracting) a wind speed $v_\mathrm{eq}$ to the receding (or approaching) pixels of the planet ring within 25$^\circ$ of latitude.

Given the above velocity field $v_\mathrm{ring}$, each pixel ($x_i,b_j$) of the planet ring contributes to the planet line profile with a Gaussian profile given by:
\begin{equation}
\mathcal{G}(\vec{v},x_i,b_j) = \mathcal{S}(x_i, b_j)~\mbox{exp}\left\{-\frac{1}{2}\frac{[\vec{v}-v_\mathrm{ring}(x_i)]^2}{\sigma^2}\right\}
\end{equation}
where the width $\sigma$ is given by 
\begin{equation}
\sigma \equiv \frac{\mbox{FWHM}_\mathrm{CRIRES}}{2\sqrt{2\log2}} \simeq 1.27~\mbox{km s}^{-1},
\end{equation}
where the FWHM of CRIRES is supposed to be constant and equal to 3~\kms.

\begin{figure}[!t]
\centering
\includegraphics[width=8cm]{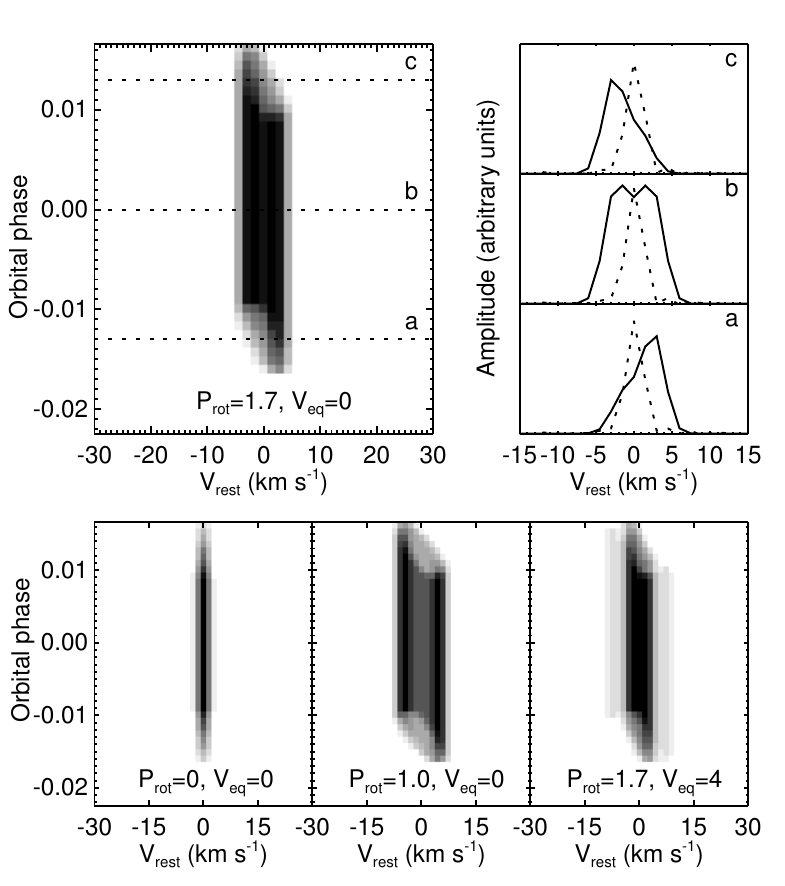}
\caption{{\sl Top-left panel:} planet broadening profiles computed for the best-fitting rotational period of \prot~=~1.7 days, as a function of planet orbital phase $\varphi$. First and fourth contact occur at $\varphi=\mp0.017$. {\sl Top-right panel:} cuts of the broadening profiles during ingress ($\varphi=-0.013$, panel $a$), at mid-transit ($\varphi=0$, panel $b$), and during egress ($\varphi=0.013$, panel $c$). For comparison, the measured instrument profile of CRIRES is shown with a dotted line. {\sl Bottom panels:}  planet broadening profiles in the case of no rotation (left), fast rotation (middle), and \prot~=~1.7 days with added equatorial super-rotation (right). In all panels, \prot\ is in days and \veq\ in \kms.}
\label{hd189_fig_profiles}
\end{figure}

Examples of the modeled line profiles as a function of orbital phase are provided in Figure~\ref{hd189_fig_profiles}. The top-right panel shows three cuts of the profiles with $P_\mathrm{rot}=1.7$~days during ingress, at mid-transit, and during egress. Distinctive features are the flattened peak at mid-transit, which eventually splits into two separate components in case of fast rotation, and the overall red-shift (or blue-shift) of the peak during ingress (or egress).

\subsection{Signal retrieval: S/N analysis}\label{snr_analysis}
For each of the CRIRES detectors, we cross correlate each of the telluric-removed spectra with the model spectra described in Section~\ref{atm_models}, Doppler shifted according to a lag vector of radial velocities in the range $\pm250$~\kms, in steps of 1.5~\kms, and convolved with the broadened line profiles computed in Section~\ref{broad_models}. As a result, we obtain the cross-correlation signal as a function of planet radial velocity, time, rotational velocity, and equatorial super-rotation velocity, that is CCF($V_\mathrm{P},t, v_\mathrm{rot}, v_\mathrm{eq}$). The total cross correlation is then computed by shifting the CCFs to the planet rest frame via linear interpolation and co-adding in time. 
Since HD~189733~b is transiting, we know the orbital inclination $i$ with very good precision. The orbital solution is consistent with zero eccentricity, and therefore we can compute the measured planet radial velocity amplitude via:
\begin{equation}\label{eq_pl_rv}
V_\mathrm{P}(t) = K_\mathrm{P}\sin[2\pi\varphi(t)] + v_\mathrm{sys} + v_\mathrm{helio}(t).
\end{equation}
The planet radial velocity amplitude \kp\ is
\begin{equation}\label{hd189_eq_kptrue}
K_\mathrm{P} = v_\mathrm{orb}\sin i,
\end{equation}
the planet orbital velocity is
\begin{equation}
v_\mathrm{orb} = \frac{2\pi a}{P_\mathrm{orb}},
\end{equation}
and the planet orbital phase $\varphi(t)$ is the fractional part of
\begin{equation}
\varphi(t) = \frac{t - T_0}{P},
\end{equation}
where $t$ is the time of our observations and $T_0$ the time of mid-transit.
By using the system parameters listed in Table~\ref{tab_syspar}, we obtain $v_\mathrm{orb} = 153.0^{+1.3}_{-1.8}$~\kms\ and \kp~$= 152.5^{+1.3}_{-1.8}$~\kms.

\begin{figure*}[!th]
\centering
\includegraphics[width=\textwidth]{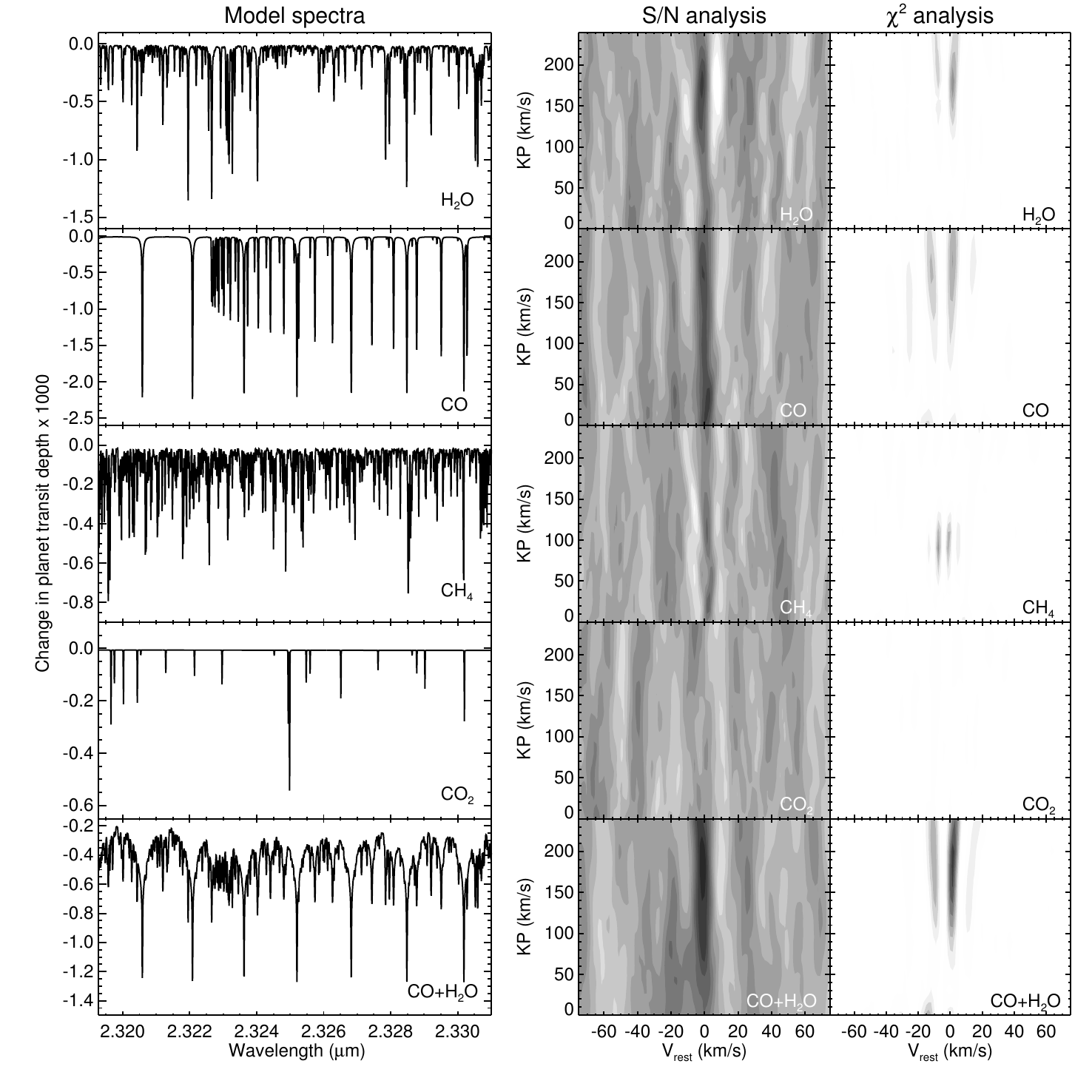}
\caption{Molecular detections in the high-resolution transmission spectrum of HD~189733~b around 2.3~\micron. The left column shows the best-fitting atmospheric model, or a model representative for the planet atmosphere when a species is not detected. The middle column shows signal-to-noise maps of the shifted and co-added cross-correlation functions. These are obtained by dividing the peak value by the standard deviation of the noise measured away from the peak. The right column show the significance maps obtained from the $\chi^2$ analysis explained in Section~\ref{chi2_analysis}. In both middle and right columns, the signal is shown as function of planet rest-frame velocity ($x$-axis) and orbital radial velocity ($y$-axis) used for shifting. From top to bottom, we plot the data for \water, CO, \methane, \cotwo, and for \water\ and CO combined. The color scheme is linear, running from $-3$ to $+7$ in S/N (middle panels) and from 0$\sigma$ to 7$\sigma$ in significance (right panels). Darker colors correspond to stronger detections. Negative S/N means anti-correlation.}
\label{fig_moldet}
\end{figure*}

While in principle we could restrict ourself to investigate the shifted and co-added CCF around the expected planet radial velocity, we also want to exclude the presence of spurious cross-correlation signal far from the planet position, and assess its impact on the planet detection. This is particularly important in this case, since stellar noise can be a significant source of spurious cross correlation (see Sections~\ref{rm_star} and \ref{onegas}). Therefore, we shift and co-add for planet rest-frame velocities between $-75$ and $+75$~\kms\ (in steps of 1.5~\kms), and planet orbital radial velocities \kp\ between 0 and 180~\kms\ (in steps of 2.25~\kms). We consequently study the total cross-correlation signal in the ($v_\mathrm{rest}, K_\mathrm{P}$) matrix, as shown in Fig.~\ref{fig_moldet}.

As in previous studies \citep{bro12, bro14, rem13, bir13}, the signal-to-noise ratio is computed by dividing the peak cross-correlation values by the standard deviation of the points away from the peak. We note, however, that this analysis has several limitations in this particular case, leading to underestimated planet signals and complicating the measurement of rotational and winds parameters. In the first place, cross correlating with model spectra with a wide range of broadening changes the level of correlation between the points of the CCF, which is sampled at a fixed resolution of 1.5~\kms. Secondly, the peak of the CCF alone cannot encode the richness of the planet signal. The planet CCF shows strong negative wings (see bottom panels of Figure~\ref{fig_moldet}). This is in part due to the autocorrelation function, and in part it is a known side effect of our strategy for removing telluric lines, already noted by \citet{sne10}. In the presence of strong planet absorption lines, the fit of the flux as a function of airmass is biased towards slightly lower values. Therefore, when dividing through the fit, we obtain excess residuals around each planet line, which generate the wings once summed in time. In addition to wings, the planet CCF shows asymmetries, autocorrelation for non-zero rest velocities, etc. All these effects are neglected if relying on the peak value of the CCF alone. Lastly, the cross correlation is a broadening operator, conceptually similar to a convolution. This means that the peak value of a CCF is naturally maximized when cross-correlating with the narrowest possible templates. This would bias our analysis to favor narrow (i.e., without rotation) planet models, affecting the determination of the planet rotational rate. A broader CCF would also reduce the precision in determining parameters uncorrelated with the width of the planet line profile, namely the maximum planet orbital velocity~\kp\ and the rest-frame velocity~\vrest. All these limitations are indeed seen and verified in these data, and we briefly describe them in Section~\ref{results} (and sub-sections). To overcome the limitations of the above analysis, we have developed a novel signal retrieval based on $\chi^2$ statistics, which we describe below.

\subsection{Signal retrieval: $\chi^2$ analysis}\label{chi2_analysis}
For each value of \kp, we cross-correlate the data with the narrow atmospheric model (i.e., without including the broadening due to planet rotation and atmospheric circulation), according to a lag vector centered on the planet radial velocity (Equation~\ref{eq_pl_rv}), and spanning $\pm90$~\kms\ around it, in steps of 1.5~\kms. Since the CCF is already centered on the planet rest-frame velocity, we obtain the total CCF by simply co-adding in time. 

We now have to define a non detection. In the case of zero planet signal and no correlated noise, we expect the cross-correlation values to be distributed as a Gaussian with mean zero. That is equivalent to fitting the CCF with a straight line and zero offset. The associated goodness of fit test requires the computation of the following $\chi^2$:
\begin{equation}
\chi^2_0(K_\mathrm{P}) = \sum_{V_\mathrm{rest}} \frac{(\mbox{CCF}_0(V_\mathrm{rest}, K_\mathrm{P})-0)^2}{s^2},
\end{equation}
where the sum is computed for $|V_\mathrm{rest}|<37.5$~\kms\ to ensure that we are only including data potentially altered by planetary signal. This corresponds to 51 cross-correlation values, or 50 degrees of freedom (d.o.f.). In the equation above $s^2$ is the variance of the CCF computed for rest-frame velocities greater than 37.5~\kms. 

We compute the p-value from $\chi^2_\mathrm{0}$, that is the probability of measuring a value higher than $\chi^2_\mathrm{0}$. It can be obtained from the Cumulative Density Function of a chi-square distribution with $n$ degrees of freedom, CDF$(x,n)$:
\begin{equation}
P_\mathrm{0} = P(x>\chi^2_\mathrm{0}) = 1 - \mbox{ CDF}(x, 50)
\end{equation}
We finally translate this probability into sigma intervals $\sigma_\mathrm{0}$ by using a Normal distribution. We use a two-tail test, since by definition the $\chi^2$ includes both negative and positive deviations around the mean. The quantity $\sigma_\mathrm{0}$ tells us how much our CCF deviates from a Gaussian distribution, regardless of the source of the signal (astrophysical or instrumental).
If the signal is originated from the atmosphere of \pname, it also embeds any broadening due to rotation and/or atmospheric circulation. However, as explained at the end of Section~\ref{snr_analysis}, we want to avoid directly cross correlating with broadened models, because that would change the statistical properties (i.e., the level of correlation) of the cross-correlation values.

Instead of cross correlating with the broadened model, we inject it at 10\% the nominal level on top of the real data, prior to removing the stellar and telluric lines. This ensures that the injected signal will be processed exactly as the real signal by our data analysis. After processing and cross-correlating the data with the narrow model, for the same choice of \kp\ and lag vectors as above, we take the difference between the CCFs of the injected and real data:
\begin{equation}
\Delta\mbox{CCF} = \mbox{CCF}_\mathrm{inj} - \mbox{CCF}_0
\end{equation}
Since the model signal was injected at a small level, the CCF noise is not significantly altered by the injection. This was tested by visually inspecting the data before and after the injection, and by comparing the noise variance, which resulted identical to within 0.1\%. Therefore, the difference $\Delta$CCF is essentially noiseless and provides a good estimator for the cross-correlation function of a real planet signal with the same parameters (\kp, \vrot, \veq) used for the injection, except for a scaling factor and (possibly) an offset. These are determined by linearly fitting $\Delta$CCF to CCF$_0$:
\begin{equation}
\mbox{CCF}_0 = c_0 + c_1\Delta\mbox{CCF} \equiv \mbox{fit}(V_\mathrm{rest},K_\mathrm{P},V_\mathrm{rot},V_\mathrm{eq})
\end{equation}
We then perform a Goodness of Fit test on our model by computing the $\chi^2$ of the residuals
\begin{eqnarray}
& \chi^2_\mathrm{model}(K_\mathrm{P}, V_\mathrm{rot}, V_\mathrm{eq}) = & \\ \nonumber
& = \sum_{V_\mathrm{rest}} \frac{\left[\mbox{CCF}_0(V_\mathrm{rest}, K_\mathrm{P})-\mbox{fit}(V_\mathrm{rest},K_\mathrm{P},V_\mathrm{rot},V_\mathrm{eq})\right]^2}{s^2}. & \\ \nonumber
\end{eqnarray}
Since we are fitting 4 parameters (\vrest, \kp, \vrot, \veq), we loose 5 d.o.f., meaning that the confidence interval will be this time
\begin{equation}
P_\mathrm{model} = P(x>\chi^2_\mathrm{model}) = 1 - \mbox{ CDF}(x, 46)
\end{equation}
We finally determine sigma intervals $\sigma_\mathrm{model}$ as before. 

The difference $\Delta\sigma = \sigma_0-\sigma_\mathrm{model}$ measures how much better our model fits the data compared to a straight line. But since a straight line means no detection by definition, $\Delta\sigma$ is also a measure of the significance of our detection. If the best-fitting model is a good match to the data, $\sigma_\mathrm{model}$ is close to zero, meaning that the residuals are perfectly consistent with Gaussian noise. Therefore $\Delta\sigma \approx \sigma_0$, as we indeed show in Section~\ref{atm_circ}. 

This approach based on $\chi^2$ has the advantage that the whole shape of the CCF, and not only its peak value, is taken into account. This means that in the presence of spurious noise no planet model signal will be a good fit to the data. Those models and the corresponding parameter set will be therefore excluded by this $\chi^2$ analysis, while they will still produce a significant signal in the S/N analysis presented in Section~\ref{snr_analysis}. This means that our $\chi^2$ analysis is also more effective at excluding spurious cross-correlation signals.


\section{Results}\label{results}

\subsection{Single-trace gases}\label{onegas}
We discuss below the signal detected from models containing \water, CO, \methane, and \cotwo\ as single-trace gases. 
These signals are shown in Figure~\ref{fig_moldet} for the S/N analysis (middle panels) and for the $\chi^2$ analysis (right panels), as a function of planet rest-frame velocity (\vrest) and radial velocity amplitude (\kp). The rotational period and equatorial velocity are set to 1.7 days and 0~\kms\ respectively, reflecting the best-fitting values discussed in Section~\ref{atm_circ}. For the S/N analysis, the error bars on the two parameters are determined by exploring the parameter space around the maximum cross-correlation signal until the S/N drops by 1. For the $\chi^2$ analysis, they correspond to the 1-$\sigma$ confidence intervals around the value with the highest significance.

\begin{itemize}
\item {\bf\water:} Water vapor is not expected to show significant absorption lines in the stellar spectrum. Therefore, a detection of water at the expected planet orbital velocity is the most reliable indicator that we are indeed measuring the planet transmission spectrum.

The total CCF from \water\ absorption lines (Figure~\ref{fig_moldet}, top panel) has a S/N~=~5.5. It peaks at \kp~=~$(183^{+38}_{-59})$~\kms, consistent with the literature value computed in Section~\ref{snr_analysis}, and \vrest = ($-1.58^{+1.65}_{-1.50}$)~\kms. From the $\chi^2$ analysis we measure a significance of 4.8$\sigma$, \kp~=~$(179^{+22}_{-21})$~\kms, and \vrest = ($-1.6\pm1.3$)~\kms.

In both analyses, no significant signal is detected at spurious orbital velocities and/or planet rest-frame velocities. This reinforces our hypothesis that no residual telluric or stellar absorption is present. It implies that the cold water vapor in the Earth's atmosphere has a high-resolution spectrum significantly different from the hot \water\ in hot-Jupiter atmospheres, as already observed in previous studies \citep{bir13,bro14}.

\item {\bf CO:} Carbon monoxide in the transmission spectrum of \pname\ is only detected when modeling and removing the stellar absorption lines and Rossiter-McLaughlin effect as explained in Section~\ref{rm_star}. If the stellar spectrum is not removed, the combination of the RM effect and the change in barycentric and stellar radial velocity produces a strong spurious signal around \kp~$\simeq$~85~\kms\ (Figure~\ref{fig_co_res}, left panel). However, when applying the S/N analysis stellar residuals at lower \kp\ are still present even after subtracting the stellar spectrum (Figure~\ref{fig_co_res}, right panel, or Figure~\ref{fig_moldet}, second middle panel from the top). As a consequence, even though the planet signal is recovered at a S/N~=~5.1 at the expected orbital velocity of \pname, the spurious stellar signal at lower \kp\ still has a comparable strength. This means that, based on the S/N analysis only, a blind search for exoplanet \pname\ based on CO alone would have failed in unambiguously identify the true source of the planet signal. The $\chi^2$ analysis instead is able to recover the CO signal unambiguously (Figure~\ref{fig_moldet}, second right panel from the top). We measure a significance of 4.1, \kp~=~$(205^{+38}_{-51})$~\kms, and \vrest~= ($-1.6^{+2.0}_{-1.8}$)~\kms. Furthermore, this signal is consistent with the \water\ detection, which is unaffected by contaminants. This demonstrates that the $\chi^2$ analysis is particularly effective in filtering out spurious cross-correlation signal not matching the expected CCF of the planet spectrum.

\begin{figure}[h]
\plotone{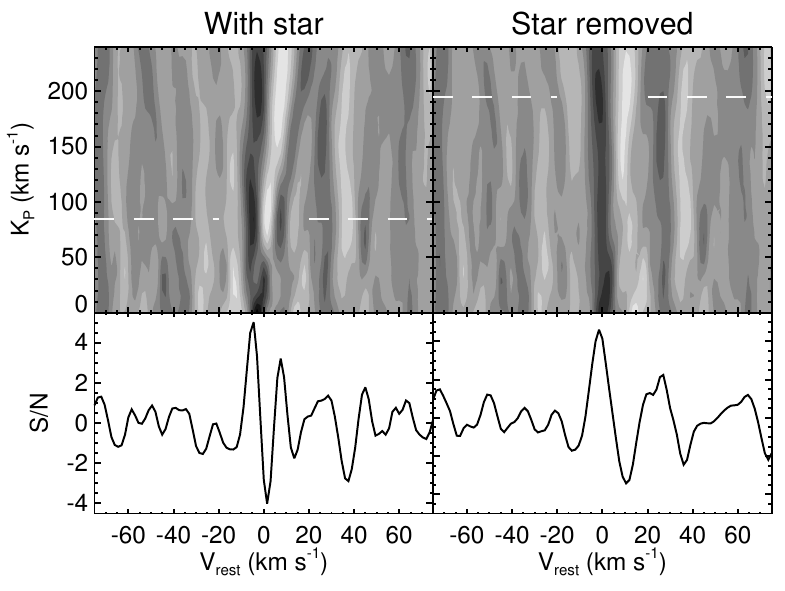}
\caption{Total cross-correlation signal when cross correlating with model containing CO only, as function of planet rest-frame velocity (\vrest) and maximum orbital radial velocity (\kp). The left panels show that if the stellar spectrum is not removed from the data (Section~\ref{rm_star}), the dominant source of signal is spurious stellar CO appearing at relatively low \kp\ (white dashed line), mostly due to the distortion of the stellar lines during the planet transit (Rossiter-McLaughlin effect). When the stellar spectrum is properly fitted and removed accounting for the shape and distortion of the spectral lines (right panels), part of the spurious signal disappears and the planet signal is detected at the expected position with a S/N of 5.2. The color scheme in the figure is linear in the range $-5 < \mbox{S/N} < 6$, with darker colors indicating higher cross-correlation values. Bottom panels show a cut of the top panels along the white dashed lines.}
\label{fig_co_res}
\end{figure} 

\item {\bf \methane:} The cross correlation with methane transmission models does not show a signal compatible with the known planet orbital velocity (Figure~\ref{fig_moldet}, third groups of panels from the top). The S/N analysis reveal a maximum S/N of 3.4 at \kp~=~98~\kms. The $\chi^2$ analysis peaks at a slightly lower \kp\ of 90~\kms, but with a higher significance of 4.6$\sigma$. This signal is inconsistent with either residual telluric \methane\ absorption (which should appear at much lower \kp), or with a genuine planet signal (expected at \kp~=~153~\kms). Residual telluric absorption is also disfavored by previous work on dayside spectra of HD~179949~b \citep{bro14}, which has shown no spurious cross-correlation from the cold methane in the Earth's atmosphere. Finally, the signal cannot be due to stellar residuals, because it does not vary in strength or position when removing the stellar lines. For the remainder of the analysis we consider \methane\ non detected as a single-trace gas.

\item {\bf \cotwo:} We do not detect a signal from carbon dioxide above the threshold of S/N~=~3. Additionally, we do not measure any significant cross correlation compatible with either telluric or stellar residuals. The absence of \cotwo\ signal was also observed in past dayside, 3.2-\micron\ observations \citep{bir13}. It is possibly due to the low number of deep \cotwo\ molecular lines in the targeted spectral range (see e.g. Figure~\ref{fig_moldet}, fourth left panel from the top), inaccuracies in the \cotwo\ high-temperature line list, or low equilibrium abundances of this molecular species, as recently suggested by \citet{hen15}.

\end{itemize}
In light of the absence of significant \cotwo\ signal and the difficult interpretation of \methane\ signals, we do not include these two species for computing mixed models. We instead mix only CO and \water\ in various relative VMRs, and discuss their signal in the next Section.

\begin{figure}[!t]
\plotone{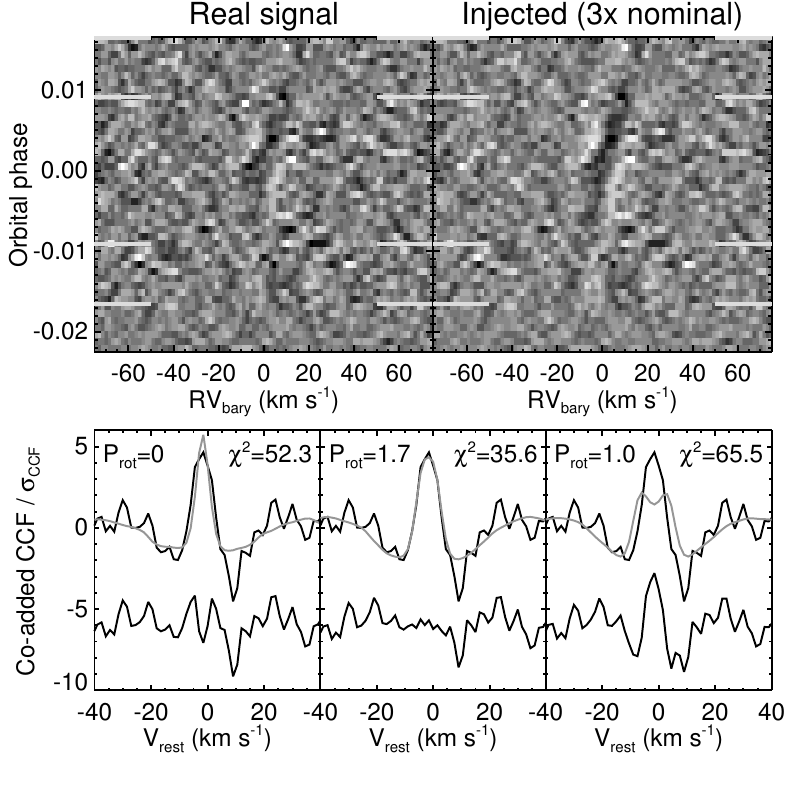}
\caption{ {\sl Top panels:} Combined absorption of \water\ and CO molecular lines in the transmission spectrum of HD~189733~b, obtained by cross correlating with the model presented in Section~\ref{best_model}. The real signal (left panel) is shown as a function of barycentric radial velocity and planet orbital phase, and compared to an artificial signal (right panel) injected at 3$\times$ the nominal level on top of the observed data. White solid lines show the first, second, third, and fourth contacts. The comparison shows that the real and the artificial signals have remarkably similar patterns in time. {\sl Bottom panels:} The total signal after shifting the cross-correlation functions to the planet rest frame and co-adding them is shown in black. Over-plotted in grey is the best-fitting model spectrum determined via $\chi^2$ analysis (Section~\ref{chi2_analysis}) for no rotation (lower-left), \prot~=~1.7 days (middle), and \prot~=~1.0 days (lower-right). The residuals are plotted at the bottom of each panel, and the corresponding $\chi^2$ is indicated at the top.}
\label{fig_trail}
\end{figure} 

\subsection{Best-fitting model and planet parameters}\label{best_model}
Among the models tested containing both CO and \water, the one that best fits our data has a cold upper atmosphere ($T_2=500$~K), and equal molecular Volume Mixing Ratio of $\log_{10}\mbox{(VMR)}=-3$ for the two species. This model maximizes the signal in both the S/N and the $\chi^2$ analysis. The total cross-correlation signal obtained with the above model, a planet rotational period of 1.7 days, and no equatorial super-rotation is shown in the bottom panel of Figure~\ref{fig_moldet}. The S/N analysis delivers S/N = 7.0, \kp~=~$(183^{+38}_{-59})$~\kms, and \vrest~=~($-0.9^{+1.8}_{-1.9}$)~\kms, while the $\chi^2$ analysis gives us a significance of 7.6$\sigma$, \kp~=~\kpval\ \kms, and \vrest~=~\vrestval~\kms. 

The CCF from the above best-fitting model is plotted as function of barycentric radial velocity and time in Figure~\ref{fig_trail} (top-left panel). For comparison, we injected the same model broadened by a rotational velocity of 3.4~\kms\ and for a planet maximum radial velocity of 153~\kms\ (Equation \ref{hd189_eq_kptrue}) on top of the observed spectra, at 3$\times$ the nominal level. After re-running the analysis and re-computing the cross correlation (top-right panel), we verify that the real and injected signals show remarkably similar patterns in both time and radial velocity. 
The two panels also show the first, second, third and fourth contacts for \pname\ (white dotted lines). Since the planet has a relatively high impact parameter, ingress and egress are relatively long compared to the total transit duration. Combined with the limb-darkened stellar disk, this causes the spectra around mid-transit to have a higher weighting than those during egress/ingress, which is well visible in the time-resolved CCF. This weighting is taken into account when co-adding the CCFs.

We now extend the $\chi^2$ analysis for this best-fitting model by varying \vrot\ and \veq, and we derive the orbital and atmospheric parameters discussed below.

\subsection{Measured planet rotation and winds}\label{atm_circ}
\begin{figure}[!ht]
\centering
\plotone{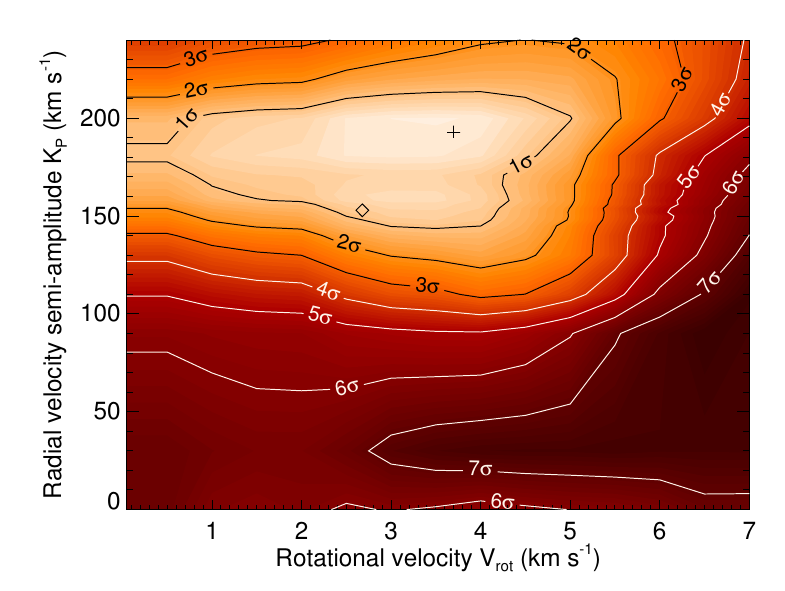}
\plotone{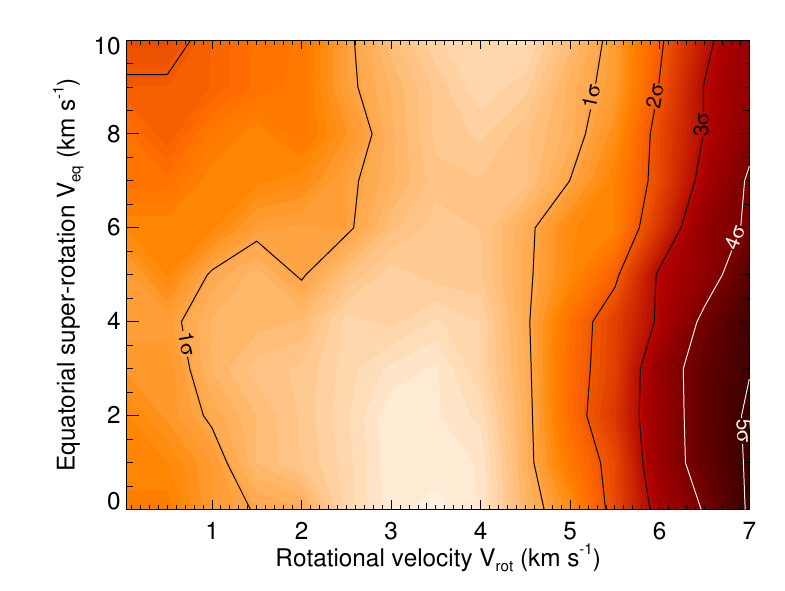}
\plotone{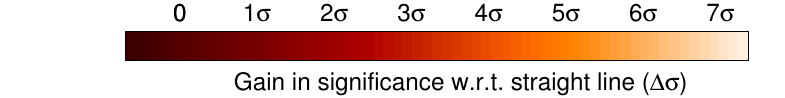}
\caption{{\sl Top panel:} Significance of the detected transmission spectrum of HD~189733~b for the best fitting model presented in Section~\ref{best_model} and varying rotational velocities \vrot\ and maximum orbital radial velocities \kp. Colored contours express the gain in significance $\Delta\sigma$ with respect to a straight-line fit to the CCF. Since a straight line means no detection by definition, $\Delta\sigma$ also represents the significance of the detection (see Sections~\ref{chi2_analysis} and \ref{atm_circ} for details). Labeled black and white contours show the confidence intervals for the two planet parameters. Their best estimate is marked with a plus sign, while the value expected by assuming tidal locking and the value of \kp\ from Equation~\ref{hd189_eq_kptrue} is indicated with a diamond. {\sl Bottom panel:} Same as above, but for planet rotational velocity \vrot\ and equatorial super-rotation velocity \veq\ at the best-fit \kp.}
\label{fig_prot_kp_veq}
\end{figure}

We computed planet broadening profiles with rotational velocities \vrot\ between 0 and 7~\kms, in steps of 0.5~\kms, corresponding to rotational periods larger than 0.8 days when adopting the planet radius in Table~\ref{tab_syspar}. We do not investigate faster rotational velocities due to the strong preference of our data for \vrot$<5.9$~\kms\ (3-$\sigma$, see discussion below and Figure~\ref{fig_prot_kp_veq}).

In addition to \vrot, we also tested equatorial super-rotation velocities \veq\ between 0 and 10~\kms, in steps of 1~\kms. The range in \veq\ is chosen based on the fact that theoretical calculations rarely show peak wind velocities exceeding 5-6~\kms\ along the terminator of a hot Jupiter, even in the absence of damping mechanisms.

Finally, we measured the rest-frame velocity \vrest\ of the peak CCF(\kp, \vrot, \veq) in order to test for global day- to night-side winds.

We present here the derived values of \kp, \vrot, \veq, and \vrest\ for our best-fitting mixed model (see Section~\ref{best_model}), obtained from the $\chi^2$ analysis described in Section~\ref{chi2_analysis}. 
When fitted and subtracted from the data, this model produces a $\chi^2_\mathrm{model}$ of 35.6. This should be compared to a $\chi^2_0$ of 168.6 when fitting a straight line. By translating these values into sigma intervals as explained is Section~\ref{chi2_analysis}, we determine that the residuals from a straight-line fit are 7.75$\sigma$ deviant from a Gaussian distribution, while those from our best-fit model are only 0.17$\sigma$ deviant. 
This means our model is favored by $\Delta\sigma = (7.75-0.17)\sigma=7.58\sigma$ compared to a straight line, which also means the planet signal is detected at the $\Delta\sigma$ confidence level, in line with previous estimates based on S/N analysis. 
The co-added CCF, the best-fit model, and the residuals are shown in Figure~\ref{fig_trail} (lower-mid panel).

We now discuss the best-fit parameters for the planet atmosphere. The error bars listed below are 1-$\sigma$ confidence intervals corresponding to a drop of 1 in $\Delta\sigma$ with respect to its maximum value.

Figure~\ref{fig_prot_kp_veq} (top panel) shows the above quantity $\Delta\sigma$ as function of rotational velocity \vrot\ and planet maximum radial velocity \kp. The planet is detected at \kp~=~\kpval~\kms. Moreover, no significant detection is present at low \kp, whereas the standard S/N analysis is affected by residual stellar noise (right panel of Figure~\ref{fig_moldet}). This demonstrates that when accounting not only for the amplitude, but also for the shape of the CCF, we can more effectively filter out spurious signals. 

We derive a planet rotational velocity of \vrot~=~\vrotval~\kms, which is consistent with the synchronous value of $v_\mathrm{sync}=(2.68\pm0.06)$~\kms, based on the known orbital period and planet radius listed in Table~\ref{tab_syspar}. Models with rotational periods longer than 1.0 days (\vrot~$<$~5.9~\kms) are strongly favored (3$\sigma$) by our analysis. So it is very unlikely that \pname\ rotates as fast as the giant planets in our solar system. However, our best-fit \vrot\ is only marginally favored (1.5$\sigma$) compared to no rotation at all. We cannot therefore confidently determine if the bulk rotation of \pname\ is perfectly synchronous, or sub-synchronous. This inability of constraining long rotational periods is a natural consequence of the CRIRES instrument resolution. At $R$=100,000, the instrumental profile under optimal conditions has a FWHM of 3~\kms, which is close to the synchronous rotational velocity. At the current S/N it becomes hard to distinguish instrumental broadening from the natural rotational broadening of the planet.

Figure~\ref{fig_prot_kp_veq} (bottom panel) shows the detection significance as function of (\vrot, \veq), and for the best-fit \kp. Our data do not show any preference for equatorial super-rotation, with the 1-$\sigma$ upper limit spanning the entire tested range.

Finally, our best-fit model broadened by the best-fit parameters peaks at \vrest~=~\vrestval\ \kms, inconsistent with strong high-altitude winds flowing from the dayside to the night-side of the planet. The implications of these measurements are discussed below. The best-fitting parameters of \pname\ are summarized in Table~\ref{tab_syspar}.

We note that the S/N analysis leads to qualitatively similar results concerning rotation and winds. It shows no preference for equatorial super-rotation, a small day-to-night side wind speed (Section~\ref{best_model}), and a preference for slow planet rotation, although the error bars on the latter are considerably bigger (\prot$>1.1$ days at 1$\sigma$). Moreover, although the S/N is maximized for \prot~=~2.0~days, longer rotational periods are totally unconstrained. This is a consequence of the fact that the cross correlation is a broadening operator. Therefore, the peak value of the cross correlation function is naturally maximized when using the narrowest possible model, which in our case is the planet model spectrum without any broadenings. Due to this bias, the S/N analysis favors longer rotational periods, which further motivates the choice of a more robust and unbiased analysis based on injection of artificial signal, their processing identical to the real data, and the comparison between the CCFs of the injected and the real data supported by $\chi^2$ statistics.


\section{Discussion}\label{discussion}
The analysis of high-dispersion transmission spectra of \pname\ around 2.3~\micron\ resulted in the detection of \water\ and CO absorption at a combined S/N of 7.0, or a significance of 7.6$\sigma$, depending on our two retrieval methods utilized. The model that best matches the shape of the observed cross-correlation function is broadened by a planet rotational velocity of \vrot~=~\vrotval~\kms, corresponding to a rotational period of \prot~=~\protval~days. We exclude at the 3-$\sigma$ level that \pname\ rotates in less than 1 day (\vrot~$>$~5.9~\kms). If the empirical (mass, rotational period) relation observed for the bodies in the solar system \citep{hug03} can be extended to extrasolar planets as suggested by \citet{sne14}, the rotation of \pname\ must have slowed considerably from its expected value in the absence of tidal effects. Our data are indeed consistent with tidal locking to the parent star (\vrot~=~2.7~\kms, \prot~=~2.2 days). However, our best-fitting model is only marginally favored (1.5$\sigma$) with respect to models with no rotational broadening. 

Beside planet bulk rotation, we were able to investigate the presence of two main patterns of atmospheric dynamics for this hot Jupiter, namely equatorial super-rotation and global day-to-night side winds along the terminator.
We were unable to constrain the former, as equatorial velocities up to 10~\kms\ in excess with respect to the planet bulk rotational velocity are equally favored in a $\chi^2$ sense. We cannot determine if this limitation is due to insufficient S/N, to a too simplistic implementation of eastward jet streams in our model (Section~\ref{broad_models}), or to a combination of both. Future efforts will be put into this matter, as low-resolution IR observations do detect a phase shift for the maximum thermal emission from the planet, meaning that the atmosphere at the sub-stellar point is significantly advected eastward by equatorial jets. We note that low- and high-resolution spectroscopy probe different pressure levels, with the former targeting the bulk thermal emission from the planet (close to the continuum level, and typically at pressures of 1.0-0.1 bar), while the latter traces the contrast between the planet radius measured in the continuum and that in the core of molecular lines, resulting in an effective pressure range of 10$^{-2}$-10$^{-3}$ bar. 

We detect a small blue-shift of the planet cross-correlation function, as the signal is maximized at a planet rest-frame velocity of \vrestval~\kms. This is only 1.4-$\sigma$ deviant from zero wind speed, but differs significantly (2.7$\sigma$) from the detection reported by \citet{wyt15} of Na absorption blue shifted by $(-8\pm 2)$~\kms. However, they probe much higher altitudes (up to 14,000 km) and lower pressures (down to 10$^{-9}$ bars) than these NIR observations. We therefore suggest that these observations target two different circulation regimes, and that the two measurements indicate the presence of a strong wind shear between the lower and upper part of the atmosphere of \pname. This also seems to be the outcome of numerical calculations in both \citet{mil12} (their Figure~4) and \citet{sho13} (their Figure~7), although a more quantitative assessment is needed, for instance by integrating the velocity field along the terminator at two different pressure levels. We note that the same optical data of \citep{wyt15} have been independently analyzed by \citet{lou15}, accounting for the effect of the Rossiter-McLaughlin effect on the stellar sodium lines. They found a blue shift of only 1.9$^{+0.7}_{-0.6}$, which is consistent with our measurement and would suggest almost no vertical wind shear.

We showed that stellar residuals, and in particular the Rossiter-McLaughlin effect, are capable of dominating, shielding and possibly contaminating the planet signal. We demonstrated that spurious signals can be filtered out to a certain extent by investigating the full shape of the cross-correlation function, and not only its peak value. For future high-dispersion observations of exoplanet atmospheres, we strongly advise to explore the shifted and co-added cross-correlation function (or more generally a shifted and co-added planet spectrum) for a range of planet orbital radial velocities, and not just at the planet rest-frame velocity. 
If a spurious signal (either telluric or stellar) is present in the data, the total cross correlation function will peak at low or zero planet orbital velocities, and at rest-frame velocities equal to the combination of the systemic velocity of the targeted system and the barycentric velocity of the observer. In that case, a modeling and subtraction of the stellar spectrum, or a better removal of telluric lines should be investigated and applied, before the results can be trusted.

We note that these transit data are less effective than previous dayside observations \citep{bro12, bro14, bir13, rem13} in constraining the amplitude of the planet orbital radial velocity (\kp), resulting in much larger error bars. This is a logical consequence of the much smaller change in planet orbital phase and radial velocity during transit. In the case of \pname\ it is further influenced by the relatively high impact parameter at which the transit occurs. This lowers the strength of the planet signal during the first and last part of the transit, which is where the planet radial velocity differs the most from zero and would therefore provide the best constraints on \kp. However, in this case we do know the planet orbital radial velocity because the orbit of \pname\ is well constrained by both stellar radial velocities and transit observations (see parameters in Table~\ref{tab_syspar}). Therefore, differently from non-transiting planets, our aim is not to precisely determine the planet motion.

In conclusion, we have demonstrated that ground-based, high-resolution spectroscopy can be successfully utilized to constrain the planet rotation and atmospheric dynamics of evolved hot Jupiters. In the future, we plan to apply more realistic velocity fields to the planet terminator, thus deriving estimates of physical quantities such as drag forces and day-night temperature contrasts from the measured wind velocities \citep[see, e.g.,][]{sho13}. 

The typical pressures probed by NIR, high-dispersion spectroscopy are intermediate between those accessible at lower resolution and those measured through high-dispersion spectroscopy in the optical. All these techniques are therefore highly complementary and should potentially be combined to get the maximum information on the vertical thermal and dynamical structure of hot-Jupiter atmosphere. By introducing chi-square analysis for retrieving signal at high spectral resolution, we hope to open up this spectral regime to powerful retrieval methods based on Bayesian analysis \citep[see, e.g.,][]{ben15}. The possibility of combining in a homogeneous fashion data at low and high spectral resolution will reinforce the synergy between space and ground observations in the JWST/E-ELT era. 

\acknowledgments

This work is based on data collected at the ESO Very Large Telescope during the DDT program 289.C-5030. We are grateful to the ESO staff at Paranal for the help in performing these observations. \\ M. Brogi acknowledges support by NASA, through Hubble Fellowship grant HST-HF2-51336 awarded by the Space Telescope Science Institute. \\ This work was performed in part under contract with the California Institute of Technology (Caltech)/Propulsion Laboratory (JPL) funded by NASA through the Sagan Fellowship Program executed by the NASA Exoplanet Science Institute. \\ This work is part of the research programs PEPSci and VICI 639.043.107, which are financed by the Netherlands Organisation for Scientific Research (NWO). \\  Funding for the Stellar Astrophysics Centre is provided by the Danish National Research Foundation (grant DNRF106). \\ We are grateful to Jean-Michel D\'esert and Catherine Huitson for their insightful comments on the latest stages of the data analysis. \\ We thank the anonymous referee the constructive and thorough reviews, which helped improving the quality of the manuscript.


\clearpage

\clearpage

\begin{table}
\begin{center}
\begin{tabular}{r|l|l}
\tableline\tableline
\multicolumn{3}{c}{Stellar parameters} \\
\tableline
$R_\mathrm{S}$ & $(0.756\pm0.018) R_\odot$ & T08 \\
$K_\mathrm{S}$ & (201.96$^{+1.07}_{-0.63}$) m s$^{-1}$ & T09 \\
\vsini & (3.316$^{+0.017}_{-0.067}$)~\kms & T09 \\
$v_\mathrm{sys}$ & $(-2.361\pm0.003)$~\kms & B05 \\
\tableline\tableline
\multicolumn{3}{c}{Orbital and transit parameters} \\
\tableline
$a$ & 0.03120(27)~AU & T09 \\
& ($8.863\pm0.020$)~\rstar & A10 \\
$P_\mathrm{orb}$ & 2.21857567(15) days & A10 \\
$T_0$ & 2,454,279.436714(15) HJD & A10 \\
$b$ & $(0.6631 \pm 0.0023)$ & A10 \\
\tableline\tableline
\multicolumn{3}{c}{Planet parameters} \\
\tableline
$R_\mathrm{P}$ & $(1.178^{+0.016}_{-0.023}) R_\mathrm{Jup}$ & T09 \\
& ($0.15531\pm0.00019$)~\rstar & A10 \\
$M_\mathrm{P}$ & ($1.138^{+0.022}_{-0.025})$~\kms & T09 \\
\kp & \kpval~\kms & This work \\
\vrot & \vrotval~\kms & This work \\
\prot\ & \protval~days & This work \\
\veq & unconstrained & This work \\
\vrest & \vrestval~\kms & This work \\
\tableline
\end{tabular}
\caption{Relevant parameters of the exoplanet system HD~189733. From top to bottom, we list stellar radius, stellar radial-velocity amplitude, stellar projected rotational velocity, systemic velocity, semi-major axis, orbital period, time of mid-transit, impact parameter, planet radius and mass, planet radial-velocity amplitude, planet rotational velocity and period, equatorial super-rotation velocity, day-to-night high-altitude wind speed. References are \citet{bou05} for B05, \citet{tor08} for T08, \citet{tri09} for T09, and \citet{ago10} for A10. \label{tab_syspar}}
\end{center}
\end{table}

\end{document}